# Photonic $p$-orbital higher-order topological insulators

Yahui Zhang[1†], Domenico Bongiovanni[1,2†], Ziteng Wang[1†], Xiangdong Wang[1†],

Shiqi Xia[1], Zhichan Hu[1], Daohong Song[1,3], Dario Jukić[4], Jingjun Xu[1],

Roberto Morandotti[2], Hrvoje Buljan[1,5], Zhigang Chen[1,3]

[1]The MOE Key Laboratory of Weak-Light Nonlinear Photonics, TEDA Applied Physics Institute and School of Physics, Nankai University, Tianjin 300457, China

[2]INRS-EMT, 1650 Blvd. Lionel-Boulet, Varennes, Quebec J3X 1S2, Canada

[3]Collaborative Innovation Center of Extreme Optics, Shanxi University, Taiyuan, Shanxi 030006, China

[4]Faculty of Civil Engineering, University of Zagreb, A. Kačića Miošića 26, 10000 Zagreb, Croatia

[5]Department of Physics, Faculty of Science, University of Zagreb, Bijenička c. 32, Zagreb 10000, Croatia

[†]These authors contributed equally to this work

**Abstract:**

The orbital degrees of freedom play a pivotal role in understanding fundamental phenomena in solid-state materials as well as exotic quantum states of matter including orbital superfluidity and topological semimetals. Despite tremendous efforts in engineering synthetic cold-atom, electronic and photonic lattices to explore orbital physics, thus far high orbitals in an important class of materials, namely, the higher-order topological insulators (HOTIs), have not been realized. Here, we demonstrate $p$-orbital corner states in a photonic HOTI, unveiling their underlying topological invariant, symmetry protection, and nonlinearity-induced dynamical rotation. In a Kagome-type HOTI, we find that topological protection of the $p$-orbital corner states demands an orbital-hopping symmetry, in addition to the generalized chiral symmetry. Due to orbital hybridization, the nontrivial topology of the $p$-orbital HOTI is "hidden" if bulk polarization is used as the topological invariant, but well manifested by the generalized winding number. Our work opens a pathway for the exploration of intriguing orbital phenomena mediated by higher band topology applicable to a broad spectrum of systems.

In condensed matter systems, an important characteristic of electrons besides charge and spin is the orbital degree of freedom (ODoF), which plays a crucial role in understanding unconventional properties in solid-state materials as well as "orbital physics" for the science and technology of correlated electrons[1]. However, due to the complexity and various degrees of freedom simultaneously involved in real materials, it has always been a challenging task to fully unravel the physics of strongly correlated electronic matter mediated by the ODoF via controlled experiments. On the other hand, the interest in synthetic orbital systems including trapped atoms in optical lattices surged rapidly over the past decade[2,3], leading to "orbital-only" quantum emulators not only for fermions but also bosons. For example, orbital lattices have been employed to demonstrate complex Bose-Einstein condensates and orbital superfluidity[4] as well as exotic topological semimetal states[5]. Clearly, the capability of purposely preparing atoms to higher orbital bands in optical lattices has opened the door for understanding orbital physics in condensed matter systems, and even for exploring new quantum states of matter not found in natural solids.

Apart from ultracold atoms in optical lattices, other synthetic platforms wherein the ODoF has emerged and caught increasing attention include semiconductor polaritonic lattices[6,7], laser-written photonic lattices[8,9], nanomechanical resonant structures[10], and atom-to-atom engineered electronic lattices[11,12]. In particular, by employing the $p$-orbital bands of polariton micropillars arranged in a honeycomb lattice, orbital edge states, and exotic Dirac dispersions were observed[13,14]. These artificial lattices give rise to the flexibility to manipulate orbital Dirac matter with transport properties and topological features difficult to achieve in conventional solid-state systems.

Using such synthetic lattice platforms, symmetry-protected topological (SPT) phases have been extensively explored due to their peculiar characteristics and unique application potential associated with robust boundary states, especially in photonics[15-18]. Notable recent endeavors have focused on the realization of higher-order topological insulators (HOTIs)[19-25] - a new class of topological materials that do not obey the conventional bulk-edge correspondence principle[18,26,27]. While their underlying physics is still an ongoing subject of investigation, HOTIs have been highly tested already for various applications, including topological nanocavities and lasers[28-30]. However, to date, all experimental studies in HOTI systems relied on the $s$-orbital band, leaving higher orbitals such as $p$-orbital HOTIs unexplored[31]. A paradigmatic model of HOTIs is based on the breathing Kagome lattices (BKLs), widely explored in the context of superconductivity[32], which exhibit the $C_3$ rotational symmetry and the generalized chiral symmetry (GCS)[33-42]. It is thus natural to ask whether one can experimentally demonstrate $p$-orbital HOTIs using a synthetic platform. In particular, what are their distinctive features and topological invariants, and which physical perspective can they open?

In this work, we experimentally demonstrate the $p$-orbital HOTIs using photonic BKLs established by a continuous-wave (CW) laser-writing technique. In a triangle-shaped nontrivial BKL, both $p_x$- and $p_y$-type orbital corner states are observed, with characteristic intensity and phase structures manifesting the zero-dimensional "zero-energy" modes. Furthermore, corner excitation leads to dynamical rotation of a dipole beam due to nonlinearity-induced lifting of $p_x$ and $p_y$ orbital mode degeneracy. We calculate the band structure and topological invariants of the orbital HOTIs and confirm their nontrivial topology from the winding number, albeit it is "hidden" in the conventional bulk polarization due to orbital coupling-induced band crossing. The SPT phase of the orbital HOTIs is found to be inherited from the lower $s$-band Hamiltonian, protected by the $C_3$ rotational symmetry, the GCS, together with a previously unknown orbital-hopping symmetry not applicable to the $s$-band HOTIs.

The photonic BKL platform used to demonstrate $p$-orbital HOTIs is illustrated in Fig. 1a, which has six-unit cells comprised of three sublattices (A, B, and C) with intracell and intercell hopping amplitudes $t_1$ and $t_2$, respectively. When $t_1 < t_2$, the BKL exhibits nontrivial topology, featuring SPT corner states as in previously studied $s$-orbital HOTIs[33-36]. In our laser-written BKL, $t_1$ and $t_2$ are tuned through respective waveguide distances, while every waveguide supports a dipole mode (see left inset, Fig. 1a), allowing for both $p_x$ and $p_y$ orbitals. When the ODoF is taken into account, additional hopping amplitudes $t_\sigma$ ($t_\pi$) must be considered, responsible for the longitudinal (transverse) hopping when the $p$-orbitals are oriented parallel (perpendicular) to the bond direction[43] (see right inset, Fig. 1a). Under the tight-binding model, the full energy spectrum is calculated from the $p$-band Hamiltonian[31] (see Method) and plotted in Fig. 1b, where localized orbital corner states in the strongly topologically nontrivial regime are marked. (Although the nontrivial phase occurs once $t_1 < t_2$ in an infinite BKL, due to the finite-size effects in an experimentally accessible BKL, localized corner states are found in a narrower window of parameters). Furthermore, the band structure for a representative nontrivial case is plotted in Fig. 1c, which displays six topological orbital corner modes in the bandgap with eigenvalues close to zero. Two typical corner modes exhibiting characteristic $p_x$- and $p_y$-type orbitals are shown in Fig. 1d1, d2: their amplitudes distribute equally among three corner sites, with no distribution in the nearest-neighbor sites but a weak amplitude in the next-nearest-neighbor (NNN) sites. If we consider, for example, the $p_x$-type orbital at the top vertex, the corner mode is exponentially localized at the A-sublattice sites with an opposite phase between dipoles in the corner and the two NNN sites (Fig. 1d1) – a characteristic feature of the HOTI corner modes[35,44]. Other orbital orientations and unequal amplitude distributions at three vertices of a finite-sized BKL are discussed in the Supplementary Material (SM).

In our experiments, we write waveguides site-by-site in a nonlinear crystal[45], establishing the topologically nontrivial (Fig. 2a1) and trivial (Fig. 2a2) BKLs. All waveguides remain intact during measurements, whereas probing of corner states is performed with an appropriately-shaped beam that can undergo linear or nonlinear propagation depending on whether a bias field is employed or not[44]. For a direct comparison of propagation through the 20-mm-long lattices, the amplitude and phase of the probe beam are pre-modulated to mimic the corner mode profiles in Fig. 1d (see SM). Measured results to illustrate orbital HOTI corner states are presented in Fig. 2b1-

f2. Specifically, for excitation of the $p_x$-type orbital corner states, the input probe beam is made of three $x$-oriented dipoles with a nonuniform intensity distribution and staggered phase, i.e., $(0, \pi)$ for the top dipole vs. $(\pi, 0)$ for the other two (Fig. 2b1), and then launched into the BKLs at the top corner (the three uppermost A-sublattice sites). After linear propagation through the trivial and nontrivial BKLs, a significant difference is observed. In the nontrivial regime, the probe beam remains to be corner-localized in the A sublattice, with no intensity reaching the neighboring B and C sites (Fig. 2c1), thus confirming the formation of orbital HOTI corner states (see Fig. 1d1). On the contrary, in the trivial regime, the same probe beam becomes strongly distorted, spreading into the edge and bulk sites with evident intensity distribution also in the B and C sites (Fig. 2c2). These results are corroborated by numerical simulations showing good agreement with the experiments (Fig. 2d1 and d2).

In another set of measurements, the input probe beam is modulated so as to excite the $p_y$-type orbital corner states as shown in Fig. 1d2, and similar results confirm the formation of a $p$-orbital mode in the orthogonal direction (see Fig. 2b2 and Fig. 2e1-f2). In addition, to validate the need for the staggered phase structure, a series of experiments and numerical simulations are performed by making the input dipoles all in phase for direct comparison (See SM). Our results show that both $p_x$- and $p_y$-type orbital corner states can be realized under appropriate excitations.

To demonstrate that the $p$-orbital BKL is indeed a symmetry-protected HOTI, we need to identify the topological invariant and prove the robustness of the corner states. The first candidate for the topological invariant is that of the $s$-band BKL: the bulk polarization[46] (see Methods), or the $Z_3$ Berry phase[31,47]. In Fig. 3a, b1, and b2, we plot the calculated polarization and 6-band structures as a function of the coupling ratio $t_\pi/t_\sigma$. Polarizations of the upper three bands are always 0, 1/3, and 2/3, but the order of these three values changes with $t_\pi/t_\sigma$ as shown in Fig. 3a. The sum of quantized polarizations of the upper three bands is always 1, however, this value must be regarded as mod 1, which implies that polarization vanishes. This suggests that either the polarization is not a good topological invariant or the $p$-orbital BKL is not an HOTI.

To resolve the above issue, we introduce an auxiliary Hamiltonian $H'(\mathbf{k},\theta)$ obtained via a unitary transformation of the $p$-band Hamiltonian $H(\mathbf{k})$[31,33], which employs a rotational operator $R(\theta)$ (see Methods and SM). The $C_3$ symmetry and the GCS are both preserved as $\theta$ is changed (see SM). At $\theta = 4\pi/3$, the auxiliary Hamiltonian is identical to the $p$-orbital BKL Hamiltonian, whereas at $\theta = 0$ it is composed of two independent BKL Hamiltonians - one for $t_\sigma$ hopping and the other for $t_\pi$ hopping, just as if we had two decoupled $s$-band BKLs (Fig. 3c). For each of these component Hamiltonians, the topological invariant is given by the bulk polarization[46], which depends solely on $t_1/t_2$. In Fig. 3d we plot the band-gap structure of the auxiliary Hamiltonian as a function of $\theta$, showing that the gap does not close as $\theta$ is varied from 0 to $4\pi/3$. Thus, the key point is that the orbital BKL described by the auxiliary Hamiltonian does not undergo a topological phase transition, so the $p$-orbital Hamiltonian at $\theta = 4\pi/3$ is topologically non-trivial, just as that for the $s$-band BKLs for $t_1/t_2 < 1$ (see SM for details).

Yet, we may ask is there a topological invariant where the nontrivial topology would not be hidden? To this end, we propose a generalized winding number $\mathcal{W}$ as defined in Methods[14,48]. Results are displayed in Fig. 3e: $\mathcal{W} = 2$ is found for the topologically nontrivial regime $(t_1/t_2 < 1)$ and $\mathcal{W} = 0$ for the trivial regime $t_1/t_2 > 1$. We emphasize that $\mathcal{W}$ is clearly a good topological invariant, independent of $\theta$. In fact, for the BKL-like orbital HOTIs, the need of using the winding number $\mathcal{W}$ is essential. This is because, for the Kagome geometry, one cannot physically decouple the system into two independent Hamiltonians - one for $t_\sigma$ and the other for $t_\pi$ hopping, akin to the chiral Hamiltonian for the quantum spin Hall effect[49,50]. Orbital hybridization leads to nontrivial band crossing and mixing from the $\sigma$ and $\pi$ subspaces as the parameter $\theta$ is changed, so one cannot relate a single band to a particular subspace in this process. The vanishing value of total bulk polarization cannot unequivocally reveal the underlying nontrivial topology. This is quite different from the 2D SSH-based HOTIs protected by the chiral symmetry[19,22-24,44,51], which can always be decomposed into two independent Hamiltonians associated with orthogonal orbital hopping (see, e.g. Ref.[51]). For the BKL-based HOTIs protected by the GCS, the use of the winding number judiciously resolves this issue. If $\theta$ changes progressively from 0 to $4\pi/3$, the band gap at zero-energy remains open as shown in Fig. 3d (or only exhibits a "trivial" touching without band inversion - see SM), thus the upper and lower three bands never entangle topologically. Since a continuous tuning of $\theta$ does not induce a phase transition, the topological invariants are entirely inherited from the decoupled auxiliary Hamiltonian at $\theta = 0$.

The case for $\mathcal{W} = 1$ in Fig. 3e merits discussion. Since the BKLs possess the $C_3$ symmetry, the winding numbers for each direction along the lattice unit vectors are identical (see Method). $\mathcal{W} = 2$ implies that both subspaces of the auxiliary Hamiltonian at $\theta = 0$ are topologically nontrivial, thus there are two SPT states (corresponding to $p_x$ and $p_y$-orbitals) in each corner. In contrast, $\mathcal{W} = 1$ indicates that only one subspace is topologically nontrivial while the other is trivial. Intuitively, one would think that half of the corner states should persist as $\theta$ is changed. However, as the trivial and nontrivial subspaces are coupled, the corner states of the whole system merge into the bulk. The absence of corner states for $\mathcal{W} = 1$ is in accord with the condition $t_{1\sigma}/t_{2\sigma} = t_{1\pi}/t_{2\pi}$ that must be held in order to pin the orbital corner states at zero-energy (see SM): the diagonal white line in Fig. 3e illustrating this condition does not cross the two $\mathcal{W} = 1$ quadrants. Thus we only have topological corner states in the $\mathcal{W} = 2$ region.

To prove the immunity of an orbital corner state against perturbations, we investigate its robustness in a rhombic BKL, which hosts only one corner state even at a finite size, by following the approach recently developed for the study of subsymmetry-protected topological states[42], neglecting the long-range hopping as a reasonable approximation. Our results (Fig. 3f1 and f2) show that orbital corner states residing in the A sublattice are protected under perturbations that respect the A-subsymmetry[42]. Moreover, we uncover that in addition to the $C_3$ symmetry and the GCS, the topological protection of the $p$-orbital HOTI corner states demands a counterintuitive *orbital-hopping symmetry* (see Method and SM). Such a symmetry demands $t_{1\sigma}/t_{2\sigma} = t_{1\pi}/t_{2\pi}$ (marked by the white dashed line in Fig. 3e), which basically implies that lattice breathing characterized by $t_1/t_2$ should not make any difference to two orthogonal orbital hopping. The lattice used in our experiments satisfies this condition approximately, as indicated by the blue cross in the region of $\mathcal{W} = 2$ in Fig. 3e.

Finally, we observe an intriguing phenomenon of dynamical rotation of orbital corner modes in nonlinear lattices. When a dipole-like beam is initially tilted away from the "equilibrium" position of the $p_x$- or $p_y$-orbital mode, corner excitation in the BKLs leads to dipole rotation towards the stable orbital orientation under self-focusing nonlinearity. A typical example is shown in Fig. 4, where both clockwise (Fig. 4b1-d1) and counter-clockwise (Fig. 4b2-d2) rotations of about 10° are observed. Such rotations can be explained by the nonlinearity-induced lifting of the orbital mode degeneracy (Fig. 4e) as detailed in SM, because nonlinearity can break the GCS and drive the corner modes away from zero-energy and even form corner solitons[41,44]

We have thus shown that $p$-orbital HOTIs in photonic BKLs have a nontrivial band topology inherited from the $s$-orbital counterparts, which is characterized by the winding number rather than the quantized bulk polarization. Higher-order orbital corner states are topologically protected by the $C_3$ symmetry, the GCS, along with the orbital-hopping symmetry not found in HOTIs where $\sigma$- and $\pi$-orbital couplings can be fully decoupled. These results have a direct impact on topological photonics involving the ODoF, now that different experimental platforms have enabled active and precise control of gain and dissipation, on-site energy, and coupling strength, in real space and synthetic dimensions[52-55]. For instance, using the ODoF, it is possible to construct a host of new topological phases including hybrid quadrupole topological insulators[56] and orbital modes in non-Euclidean surfaces with disclinations[57]. Our results may prove relevant to those studies, and other orbital systems beyond photonics.

## Methods

**Orbital Hamiltonian and the winding number:**

Under the tight-binding approximation, the $p$-orbital corner states in a BKL can be found from the real-space Hamiltonian $H$ expressed as[31]:

$$H = \sum_{m,n} t_{1\sigma} \left( a_{1_{m,n}}^\dagger b_{1_{m,n}} + b_{2_{m,n}}^\dagger c_{2_{m,n}} + c_{3_{m,n}}^\dagger a_{3_{m,n}} \right) + \sum_{m,n} t_{2\sigma} \left( a_{1_{m,n}}^\dagger b_{1_{m+1,n}} + b_{2_{m,n}}^\dagger c_{2_{m-1,n+1}} + c_{3_{m,n}}^\dagger a_{3_{m,n+1}} \right) + \sum_{m,n} t_{1\pi} \left( \tilde{a}_{1_{m,n}}^\dagger \tilde{b}_{1_{m,n}} + \tilde{b}_{2_{m,n}}^\dagger \tilde{c}_{2_{m,n}} + \tilde{c}_{3_{m,n}}^\dagger \tilde{a}_{3_{m,n}} \right) + \sum_{m,n} t_{2\pi} \left( \tilde{a}_{1_{m,n}}^\dagger \tilde{b}_{1_{m+1,n}} + \tilde{b}_{2_{m,n}}^\dagger \tilde{c}_{2_{m-1,n+1}} + \tilde{c}_{3_{m,n}}^\dagger \tilde{a}_{3_{m,n+1}} \right) + h.c., \tag{M1}$$

where $a_{i_{m,n}}$ ($\tilde{a}_{i_{m,n}}$), $b_{i_{m,n}}$ ($\tilde{b}_{i_{m,n}}$) and $c_{i_{m,n}}$ ($\tilde{c}_{i_{m,n}}$) ($i = 1,2,3$) are the $\sigma$-type ($\pi$-type) annihilation operators at A, B, and C sublattice sites in the $(m,n)$th unit cell, directed along primitive lattice vectors $\boldsymbol{e}_i$. The coefficients $t_{1\sigma}$ ($t_{2\sigma}$) and $t_{1\pi}$ ($t_{2\pi}$) are the intracell (intercell) orbital coupling strengths (see Fig. 1a, and SM). In Fig. 1, the linear band structure ($\beta_L$) is obtained by diagonalizing $H$ for different dimerization parameters $t_1/t_2$ and orbital coupling ratios $t_{1\sigma}/t_{2\sigma} = t_{1\pi}/t_{2\pi}$. The $p$-orbital mode distribution is obtained by retrieving both $p_x$- and $p_y$-components from the calculated eigenvectors of $H$. In momentum space, the Fourier transform of $H$ for an orthogonal basis along $x$- and $y$-directions reads as

$$H(\boldsymbol{k}) = \begin{pmatrix} 0 & H_1^\dagger & H_3 \\ H_1 & 0 & H_2^\dagger \\ H_3^\dagger & H_2 & 0 \end{pmatrix}, \tag{M2}$$

where the matrix entries $H_i$ ($i = 1,2,3$) are $2 \times 2$ matrices defined as $H_i(\boldsymbol{k}) = \boldsymbol{e}_i \boldsymbol{e}_i^\dagger \left( t_{1\sigma} + t_{2\sigma} e^{i\boldsymbol{k}\cdot\boldsymbol{e}_i} \right) + \boldsymbol{d}_i \boldsymbol{d}_i^\dagger \left( t_{1\pi} + t_{2\pi} e^{i\boldsymbol{k}\cdot\boldsymbol{e}_i} \right)$, with $\boldsymbol{k} = (k_x, k_y)$ being the transverse wave vector, and $\boldsymbol{d}_i$ the unit vector orthogonal to $\boldsymbol{e}_i$. The SPT phase of the $p$-band BKL model is identified from an auxiliary Hamiltonian $H'(\boldsymbol{k}, \theta)$, which is obtained by applying a unitary transformation on $H(\boldsymbol{k})$ (see SM for details). The structure of $H'(\boldsymbol{k}, \theta)$ is identical to that in Eq. (M2), with the matrix entries $H_i(\boldsymbol{k})$ replaced by $H_i'(\boldsymbol{k}, \theta) = \begin{pmatrix} t_{1\sigma} + t_{2\sigma} e^{i\boldsymbol{k}\cdot\boldsymbol{e}_i} & 0 \\ 0 & t_{1\pi} + t_{2\pi} e^{i\boldsymbol{k}\cdot\boldsymbol{e}_i} \end{pmatrix} R(\theta)$, where $R(\theta) = \begin{pmatrix} \cos\theta & \sin\theta \\ -\sin\theta & \cos\theta \end{pmatrix}$ is the rotation operator. The orbital Kagome Hamiltonian $H(\boldsymbol{k})$ (M2) is fully equivalent to $H'(\boldsymbol{k}, \theta = 4\pi/3)$, i.e., at $\theta = 4\pi/3$. However, in our analysis, $R(\theta)$ is used to artificially tune the auxiliary Hamiltonian $H'(\boldsymbol{k}, \theta)$ from the orbital-decoupled $\theta = 0$ to the physical $\theta = 4\pi/3$ Kagome model, to unequivocally identify the topological phase.

The winding number coefficients $\mathcal{W}_i$ are calculated from the matrix elements of $H'(\mathbf{k}, \theta)$ along each direction pointed by $\mathbf{e}_i$ in the first Brillouin zone through the following expression:

$$\mathcal{W}_i = \frac{1}{2\pi} \int_0^{2\pi} dk_i \frac{d\Phi_i(k_i)}{dk_i}, \quad i = 1,2,3 \tag{M3}$$

where $\det[H'_i(\mathbf{k}, \theta)] = |\det[H'_i(\mathbf{k}, \theta)]| e^{i\Phi_i(\mathbf{k},\theta)}$, with $\Phi_i(\mathbf{k}, \theta)$ being the argument of $\det[H'_i(\mathbf{k}, \theta)]$, and $k_i$ the wave vectors along the direction $\mathbf{e}_i$. In general, for an arbitrary value of $\theta$, we have $\mathcal{W} = \mathcal{W}_1 = \mathcal{W}_2 = \mathcal{W}_3$, i.e., are all $\mathcal{W}_i$ are equal as long as the $C_3$ symmetry is preserved in the system.

**Continuous model and numerical simulations:**

To provide a direct correspondence to the experimental results in our photonic platform based on a saturable photorefractive (PR) crystal (SBN:61), numerical simulations from the following continuous-model nonlinear Schrödinger-like equation (NLSE) are compared to the measurements[44]:

$$i\frac{\partial \Psi}{\partial z} = -\frac{1}{2k} \nabla_\perp^2 \Psi - \frac{k\Delta n}{n_0} \frac{\Psi}{1 + I_L(x,y) + I_P(x,y)}. \tag{M4}$$

Here, $\Psi(x, y, z)$ is the electric field envelope, where $x$ and $y$ are the transverse coordinates and $z$ the propagation distance, $k = 2\pi n_0/\lambda_0$ is the wavenumber in the medium, with $n_0 = 2.35$ being the crystal refractive index and $\lambda_0 = 532$ nm the laser wavelength. $\Delta n$ is the refractive index change determined by the crystal electro-optic coefficient and the bias field. $I_L(x, y)$ and $I_P(x, y)$ are the intensity patterns of the lattice-writing and probing beams, respectively. The strength of the nonlinearity is controlled by the probe-beam intensity and the bias field[44]. Solutions to Eq. (M4) for orbital corner states in both trivial and nontrivial BKL geometries are found via the split-step Fourier transform method.

**Experimental Method:**

Experimental BKLs in trivial and nontrivial geometries are established by using an ordinarily-polarized CW green laser and a site-to-site writing process of each component waveguide[45]. To attain a variable planar distribution, a spatial light modulator (SLM) is also employed to modulate the initial phase of the laser beam in the Fourier space. The effective PR "memory effect" guarantees that all waveguides remain intact during the measurement time window. The excitation of the orbital corner states is made with a low-power extraordinarily-polarized probe beam, which undergoes linear propagation through the lattices for the results presented in Fig. 2. The probe beam is appropriately shaped by amplitude and phase modulation in order to excite the $p$-orbital corner modes (see SM). In the nonlinear regime, a tilted dipole-like probe beam is launched at the top corner of the nontrivial BKL, and the strength of the self-focusing nonlinearity can be controlled by the probe-beam intensity and the bias field.

**Acknowledgments**

We acknowledge financial support from the National Key R&D Program of China (2017YFA0303800), the National Natural Science Foundation (12134006, 11922408), and the QuantiXLie Center of Excellence, a project co-financed by the Croatian Government and the European Union through the European Regional Development Fund the Competitiveness and Cohesion Operational Programme (KK.01.1.1.01.0004). D. B. acknowledges support from the 66 Postdoctoral Science Grant of China. R.M. acknowledge support from the NSERC Discovery Grant and Canada Research Chair Programs.

**References:**

1. Tokura, Y.,& Nagaosa, N. Orbital Physics in Transition-Metal Oxides. *Science* **288**, 462-468 (2000).
2. Lewenstein, M.,& Liu, W. V. Orbital Dance. *Nat. Phys.* **7**, 101-103 (2011).
3. Li, X.,& Liu, W. V. Physics of Higher Orbital Bands in Optical Lattices: A Review. *Rep. Prog. Phys.* **79**, 116401 (2016).
4. Wirth, G., Ölschläger, M.,& Hemmerich, A. Evidence for Orbital Superfluidity in the P-Band of a Bipartite Optical Square Lattice. *Nat. Phys.* **7**, 147-153 (2010).
5. Sun, K. *et al.* Topological Semimetal in a Fermionic Optical Lattice. *Nat. Phys.* **8**, 67-70 (2011).
6. Jacqmin, T. *et al.* Direct Observation of Dirac Cones and a Flatband in a Honeycomb Lattice for Polaritons. *Phys. Rev. Lett.* **112**, 116402 (2014).
7. St-Jean, P. *et al.* Lasing in Topological Edge States of a One-Dimensional Lattice. *Nat. Photon.* **11**, 651-656 (2017).
8. Cáceres-Aravena, G., Torres, L. E. F. F.,& Vicencio, R. A. Topological and Flat-Band States Induced by Hybridized Linear Interactions in One-Dimensional Photonic Lattices. *Phys. Rev. A* **102**, 023505 (2020).
9. Guzman-Silva, D., Caceres-Aravena, G.,& Vicencio, R. A. Experimental Observation of Interorbital Coupling. *Phys. Rev. Lett.* **127**, 066601 (2021).
10. Ma, J. *et al.* Nanomechanical Topological Insulators with an Auxiliary Orbital Degree of Freedom. *Nat. Nanotechnol.* **16**, 576-583 (2021).
11. Slot, M. R. *et al.* Experimental Realization and Characterization of an Electronic Lieb Lattice. *Nat. Phys.* **13**, 672-676 (2017).
12. Slot, M. R. *et al.* P-Band Engineering in Artificial Electronic Lattices. *Phys. Rev. X* **9**, 011009 (2019).
13. Milićević, M. *et al.* Type-III and Tilted Dirac Cones Emerging from Flat Bands in Photonic Orbital Graphene. *Phys. Rev. X* **9**, 031010 (2019).
14. Milićević, M. *et al.* Orbital Edge States in a Photonic Honeycomb Lattice. *Phys. Rev. Lett.* **118**, 107403 (2017).
15. Rechtsman, M. C. *et al.* Photonic Floquet Topological Insulators. *Nature* **496**, 196-200 (2013).
16. Hafezi, M. *et al.* Imaging Topological Edge States in Silicon Photonics. *Nat. Photon.* **7**, 1001-1005 (2013).
17. Khanikaev, A. B. *et al.* Photonic Topological Insulators. *Nat. Mater.* **12**, 233-239 (2013).
18. Ozawa, T. *et al.* Topological Photonics. *Rev. Mod. Phys.* **91**, 015006 (2019).
19. Benalcazar, W. A., Bernevig, B. A.,& Hughes, T. L. Quantized Electric Multipole Insulators. *Science* **357**, 61-66 (2017).
20. Song, Z., Fang, Z.,& Fang, C. (d-2)-Dimensional Edge States of Rotation Symmetry Protected Topological States. *Phys. Rev. Lett.* **119**, 246402 (2017).
21. Schindler, F. *et al.* Higher-Order Topological Insulators. *Sci. Adv.* **4**, eaat0346 (2018).
22. Serra-Garcia, M. *et al.* Observation of a Phononic Quadrupole Topological Insulator. *Nature* **555**, 342-345 (2018).
23. Peterson, C. W. *et al.* A Quantized Microwave Quadrupole Insulator with Topologically Protected Corner States. *Nature* **555**, 346-350 (2018).

24. Imhof, S. *et al.* Topolectrical-Circuit Realization of Topological Corner Modes. *Nat. Phys.* **14**, 925-929 (2018).
25. Qi, Y. *et al.* Acoustic Realization of Quadrupole Topological Insulators. *Phys. Rev. Lett.* **124**, 206601 (2020).
26. Kim, M., Jacob, Z., & Rho, J. Recent Advances in 2D, 3D, and Higher-Order Topological Photonics. *Light Sci. Appl.* **9**, 130 (2020).
27. Xie, B. *et al.* Higher-Order Band Topology. *Nat. Rev. Phys.* **3**, 520-532 (2021).
28. Ota, Y. *et al.* Photonic Crystal Nanocavity Based on a Topological Corner State. *Optica* **6**, 786-789 (2019).
29. Zhang, W. *et al.* Low-Threshold Topological Nanolasers Based on the Second-Order Corner State. *Light Sci. Appl.* **9**, 109 (2020).
30. Kim, H. R. *et al.* Multipolar Lasing Modes from Topological Corner States. *Nat. Commun.* **11**, 5758 (2020).
31. Lu, X., Chen, Y., & Chen, H. Orbital Corner States on Breathing Kagome Lattices. *Phys. Rev. B* **101**, 195143 (2020).
32. Neupert, T. *et al.* Charge Order and Superconductivity in Kagome Materials. *Nat. Phys.* **18**, 137-143 (2021).
33. Ezawa, M. Higher-Order Topological Insulators and Semimetals on the Breathing Kagome and Pyrochlore Lattices. *Phys. Rev. Lett.* **120**, 026801 (2018).
34. El Hassan, A. *et al.* Corner States of Light in Photonic Waveguides. *Nat. Photon.* **13**, 697-700 (2019).
35. Ni, X. *et al.* Observation of Higher-Order Topological Acoustic States Protected by Generalized Chiral Symmetry. *Nat. Mater.* **18**, 113-120 (2019).
36. Xue, H. *et al.* Acoustic Higher-Order Topological Insulator on a Kagome Lattice. *Nat. Mater.* **18**, 108-112 (2019).
37. Ezawa, M. Higher-Order Topological Electric Circuits and Topological Corner Resonance on the Breathing Kagome and Pyrochlore Lattices. *Phys. Rev. B* **98**, 201402(R) (2018).
38. Kempkes, S. N. *et al.* Robust Zero-Energy Modes in an Electronic Higher-Order Topological Insulator. *Nat. Mater.* **18**, 1292-1297 (2019).
39. Li, M. *et al.* Higher-Order Topological States in Photonic Kagome Crystals with Long-Range Interactions. *Nat. Photon.* **14**, 89-94 (2019).
40. Peterson Christopher, W. *et al.* A Fractional Corner Anomaly Reveals Higher-Order Topology. *Science* **368**, 1114-1118 (2020).
41. Kirsch, M. S. *et al.* Nonlinear Second-Order Photonic Topological Insulators. *Nat. Phys.* **17**, 995-1000 (2021).
42. Wang, Z. *et al.* Sub-Symmetry Protected Topological States. arXiv:2205.07285 [physics.optics] (2022).
43. Liu, W. V., & Wu, C. Atomic Matter of Nonzero-Momentum Bose-Einstein Condensation and Orbital Current Order. *Phys. Rev. A* **74**, 013607 (2006).
44. Hu, Z. *et al.* Nonlinear Control of Photonic Higher-Order Topological Bound States in the Continuum. *Light Sci. Appl.* **10**, 164 (2021).
45. Xia, S. *et al.* Unconventional Flatband Line States in Photonic Lieb Lattices. *Phys. Rev. Lett.* **121**, 263902 (2018).

46. Benalcazar, W. A., Li, T., & Hughes, T. L. Quantization of Fractional Corner Charge in Cn-Symmetric Higher-Order Topological Crystalline Insulators. *Phys. Rev. B* **99**, 245151 (2019).
47. Wakao, H. *et al.* Higher-Order Topological Phases in a Spring-Mass Model on a Breathing Kagome Lattice. *Phys. Rev. B* **101**, 094107 (2020).
48. Kane, C. L., & Lubensky, T. C. Topological Boundary Modes in Isostatic Lattices. *Nat. Phys.* **10**, 39-45 (2013).
49. Hasan, M. Z., & Kane, C. L. Colloquium: Topological Insulators. *Rev. Mod. Phys.* **82**, 3045-3067 (2010).
50. Qi, X.-L., & Zhang, S.-C. Topological Insulators and Superconductors. *Rev. Mod. Phys.* **83**, 1057-1110 (2011).
51. Pelegrí, G. *et al.* Second-Order Topological Corner States with Ultracold Atoms Carrying Orbital Angular Momentum in Optical Lattices. *Phys. Rev. B* **100**, 205109 (2019).
52. Xia, S. *et al.* Nonlinear Tuning of PT Symmetry and Non-Hermitian Topological States. *Science* **372**, 72-76 (2021).
53. Pernet, N. *et al.* Gap Solitons in a One-Dimensional Driven-Dissipative Topological Lattice. *Nat. Phys.* **18**, 678-684 (2022).
54. Pourbeyram, H. *et al.* Direct Observations of Thermalization to a Rayleigh–Jeans Distribution in Multimode Optical Fibres. *Nat. Phys.* **18**, 685-690 (2022).
55. Leefmans, C. *et al.* Topological Dissipation in a Time-Multiplexed Photonic Resonator Network. *Nat. Phys.* **18**, 442-449 (2022).
56. Schulz, J. *et al.* Photonic Quadrupole Topological Insulator Using Orbital-Induced Synthetic Flux. arXiv:2204.06238 [physics.optics] (2022).
57. Chen, Y. *et al.* P-Orbital Disclination States in Non-Euclidean Geometries. arXiv:2201.10039 [cond-mat.mes-hall] (2022).

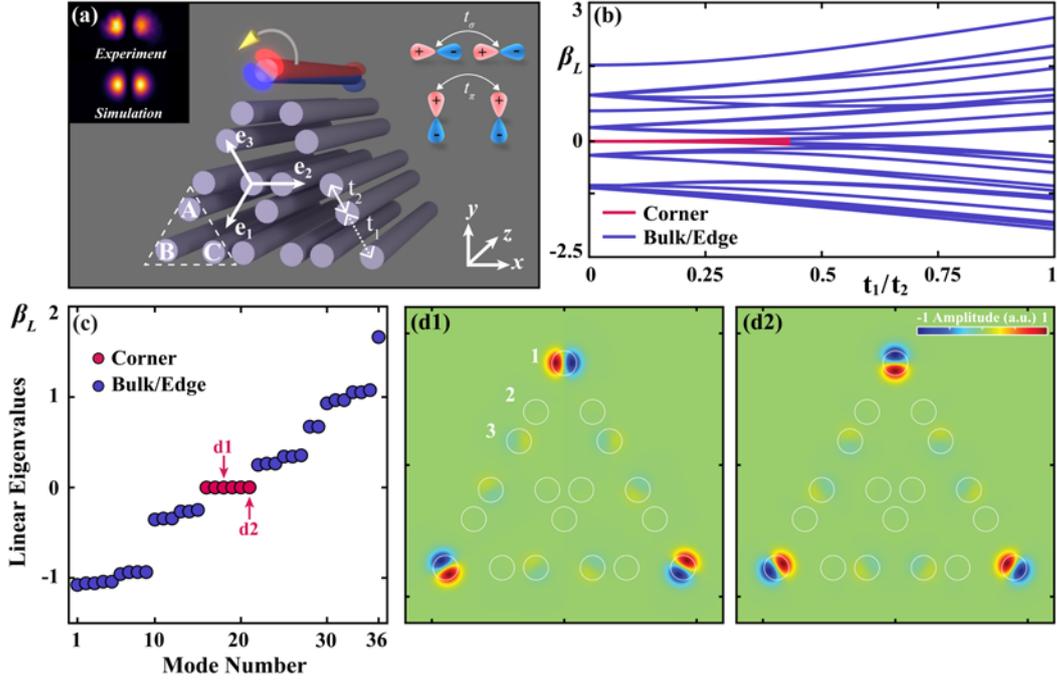

**Fig. 1: Illustration of "zero-energy" $p$-orbital topological corner states in a BKL.** (a) Schematic diagram of the triangular BKL with six-unit cells, where the white dashed triangle marks one unit cell consisting of three sublattices (A, B, and C), and $t_1$ and $t_2$ indicate the intracell and intercell hopping amplitudes, respectively. Each lattice site corresponds to a laser-written waveguide, which supports $p$-orbital modes (see the left inset). Two types of orbital hopping amplitudes, $t_\sigma$ and $t_\pi$, are illustrated in the right inset. In the top vertex of the lattice, we sketch the excitation and rotation of an orbital corner state under nonlinear propagation. (b) Calculated eigenvalue spectrum $\beta_L$ as a function of $t_1/t_2$ at $t_\pi/t_\sigma = -0.3$ for the finite-size BKL shown in (a), where distinct "zero-energy" corner states in the highly nontrivial regime are marked in red. (c) Calculated band structure for the BKL with $t_1/t_2 = 0.1$, showing six corner states (red dots) at $\beta_L \approx 0$, where two representatives ($p_x$- and $p_y$-type modes relative to the top corner) are plotted in (d1, d2). Notice the characteristic distribution of a topological $p$-orbital corner mode: e.g., for the $p_x$-mode in (d1), it is localized mainly at the corner site 1, with a staggered phase distribution in the NNN site 3 but zero amplitude at the nearest-neighbor site 2.

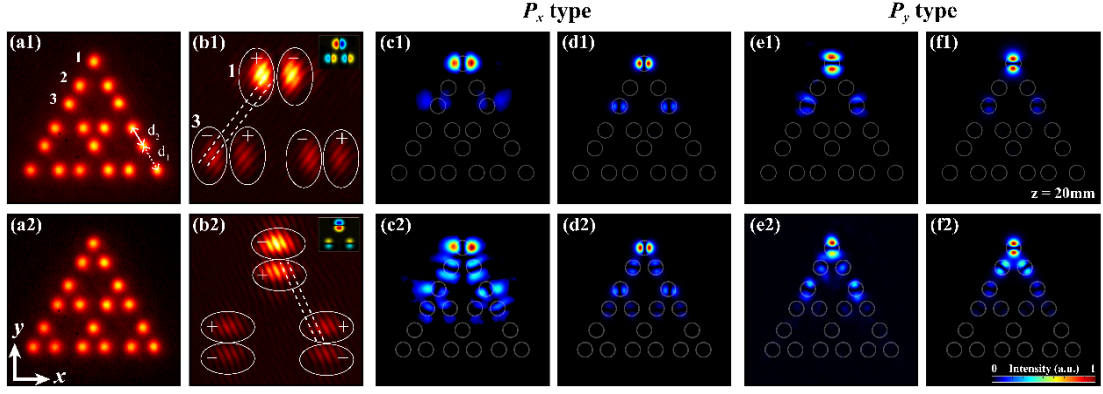

**Fig. 2**: **Demonstration of *p*-orbital topological corner states in a photonic BKL.** (a1, a2) Experimental BKLs in nontrivial (a1) and trivial (a2) regimes. (b1, b2) Zoom-in interferograms of the probe beam with a reference plane wave, showing the out-of-phase relation between corner site 1 and its NNN site 3 for both $p_x$- (b1) and $p_y$-type (b2) orbital excitations. The white ellipses outline the three-dipole probe beams, where plus/minus signs represent $0/\pi$ phase for each dipole, and dashed straight lines are added to see the phase structure from shifted (b1) and aligned (b2) interference fringes, confirming the staggered phase structures (see also numerical results in right insets). (c1, c2) Observed output intensity patterns in nontrivial (c1) and trivial (c2) BKLs stemming from a $p_x$-orbital corner excitation. (d1, d2) Numerical simulations corresponding to (c1, c2). Results from the $p_y$-orbital corner excitation are shown in (e1, e2) and (f1, f2). Grey circles in (c1-f2) mark the BKL sites. (Lattice parameters are $d_1 = 39$ μm and $d_2 = 31$μm for $p_x$-orbital excitation in the nontrivial BKL).

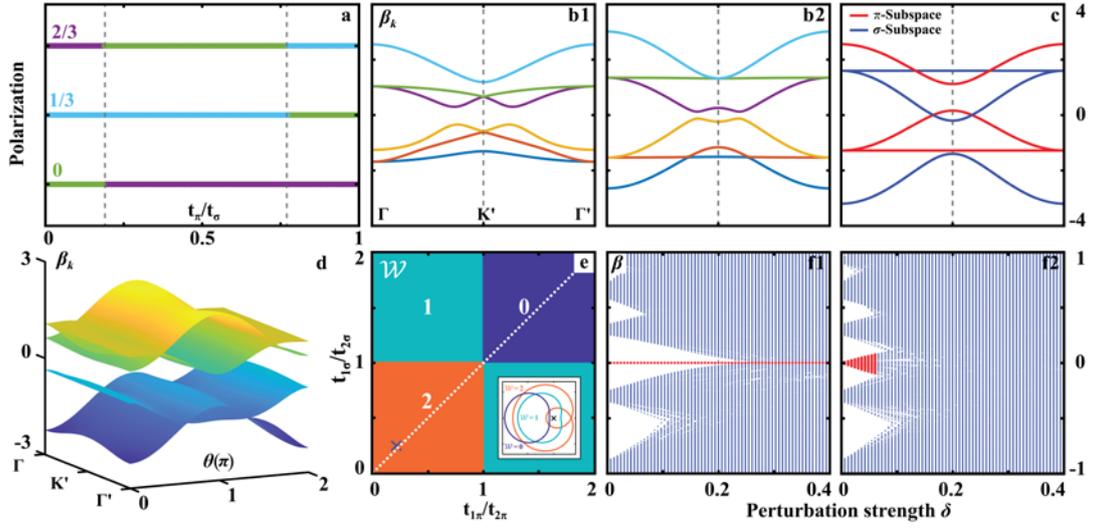

**Fig. 3: Theoretical analysis of the $p$-orbital HOTI: topological invariant and corner-mode robustness.** (a) Bulk polarization of the upper three bands (blue, green, and purple as plotted in b1 and b2) as a function of the orbital coupling ratio $t_\pi/t_\sigma$, exhibiting step jumps at band-crossing points (marked by vertical dashed lines). The sum of the quantized polarizations of the upper three bands is $(0 + 1/3 + 2/3)\ mod\ 1 = 0$. (b1-b2) Band structures plotted for (b1) $t_\pi/t_\sigma = -0.19$ and (b2) $t_\pi/t_\sigma = -0.77$ at $t_1/t_2 = 0.6$ and $\theta = 4\pi/3$, showing the change in band-crossing of the upper three bands. (c) Band structure of the auxiliary Hamiltonian $H'(\boldsymbol{k}, \theta)$ calculated for $t_1/t_2 = 0.6$ and $t_\pi/t_\sigma = -0.8$ at $\theta = 0$, which is equivalent to two sets of decoupled $s$-band BKLs: red for one set and dark blue for the other set. (d) 3D plot of the band structure of $H'(\boldsymbol{k}, \theta)$ as a function of $\theta$, showing that the gap at zero energy remains open for any $\theta$ (or any orbital hybridization). (e) Calculated winding numbers, where the white dotted line marks the orbital-hopping condition $t_{1\sigma}/t_{2\sigma} = t_{1\pi}/t_{2\pi}$ required for topological protection of the orbital corner states, circles in the lower inset illustrate distinct windings for each case, and the blue cross in the $\mathcal{W} = 2$ region corresponds to the experimental parameters with which the $p_x$-orbital HOTI is realized. (f1-f2) Robustness test of orbital corner states in a rhombic BKL attained by applying random perturbations with increasing strengths $\delta$ for A-subsymmetry-preserving (f1) and -breaking (f2) cases. The corner modes (red circles) remain at zero-energy in (f1).

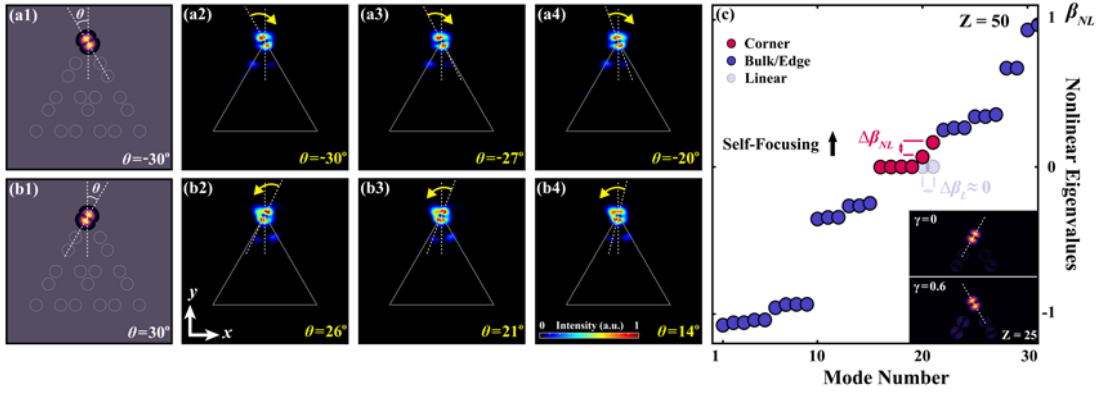

**Fig. 4**: **Nonlinearity-induced dynamical rotation of $p$-orbital corner states.** (a1, b1) Excitation of the top corner of nontrivial BKL with a dipole-like beam initially tilted at (a1) $\theta = -30°$ and (b1) $\theta = 30°$ with respect to the $y$-axis. (a2-a4) Experimental output of the orbital corner state, showing a clockwise rotation under the action of a self-focusing nonlinearity (with a bias field of 175 kV/m). (b2-b4) Opposite rotation, observed under the same nonlinearity when the initial excitation corresponds to (b1). Grey triangle in each panel outlines the BKL boundaries. Note that no rotation occurs when the dipole is initially oriented along the $y$-axis. (c) Orbital band structure calculated from the discrete NLSE at dimensionless distance $Z = 50$ for an incident probe beam with $\theta = 30°$ propagating under linear ($\gamma = 0$) and nonlinear ($\gamma = 0.6$) conditions. A tilted dipole-like beam excites two orthogonal orbital corner modes which are nearly degenerate in the linear regime ($\Delta\beta_L \approx 0$), but the nonlinearity lifts the degeneracy and increases the difference in nonlinear eigenvalues $\Delta\beta_{NL}$, leading to rotation of the orbital state. Insets illustrate the orientation of a probe beam at $Z = 25$ for both linear and nonlinear evolutions.